\documentclass[twocolumn]{article}
\usepackage[margin=1in]{geometry}
\usepackage{multirow}
\usepackage{graphicx}% Include figure files
\usepackage{dcolumn}% Align table columns on decimal point
\usepackage{bm}% bold math

\usepackage{nameref}
\usepackage{lipsum}% http://ctan.org/pkg/lipsum
\usepackage{multirow}
\usepackage{graphicx}% Include figure files
\usepackage{dcolumn}% Align table columns on decimal point
\usepackage{bm}% bold math
\usepackage{newtxtext}
\usepackage[utf8]{inputenc}
\usepackage[T1]{fontenc}
\usepackage{mathtools}
\usepackage{mathrsfs}
\usepackage{amsmath, amssymb}
\usepackage[dvipsnames]{xcolor}
\usepackage{color}

\usepackage[utf8]{inputenc}
\usepackage[T1]{fontenc}
\usepackage{mathptmx}

\usepackage{mathtools}
\usepackage{ mathrsfs }

\usepackage{amsmath, amssymb}
\usepackage{authblk}

\usepackage[dvipsnames]{xcolor}
\usepackage{color}
\usepackage{hyperref}
\hypersetup{colorlinks=true,
            linkcolor = NavyBlue,
            citecolor = NavyBlue,
            urlcolor = NavyBlue}
\usepackage[capitalize]{cleveref}

\graphicspath{{./Figures/}}
\title{Dynamic State Analysis of a Driven Magnetic Pendulum using Ordinal Partition Networks and Topological Data Analysis}
\author[1]{Audun Myers}
\author[2]{Firas A.~Khasawneh}

\affil[1,2]{Department of Mechanical Engineering, Michigan State University.}

\affil[1]{Email: myersau3@msu.edu}

\date{}

\begin{document}

\maketitle    

%%%%%%%%%%%%%%%%%%%%%%%%%%%%%%%%%%%%%%%%%%%%%%%%%%%%%%%%%%%%%%%%%%%%%%
\begin{abstract}
  %!TEX root = ..\MSP_main_IDETC.tex
%-------------------------------
%*******************************
The use of complex networks for time series analysis has recently shown to be useful as a tool for detecting dynamic state changes for a wide variety of applications. 
In this work, we implement the commonly used ordinal partition network to transform a time series into a network for detecting these state changes for the simple magnetic pendulum. 
The time series that we used are obtained experimentally from a base-excited magnetic pendulum apparatus, and numerically from the corresponding governing equations.
The magnetic pendulum provides a relatively simple, non-linear example demonstrating transitions from periodic to chaotic motion with the variation of system parameters. 
For our method, we implement persistent homology, a shape measuring tool from Topological Data Analysis (TDA), to summarize the shape of the resulting ordinal partition networks as a tool for detecting state changes. 
We show that this network analysis tool provides a clear distinction between periodic and chaotic time series. 
Another contribution of this work is the successful application of the networks-TDA pipeline, for the first time, to signals from non-autonomous nonlinear systems. 
This opens the door for our approach to be used as an automatic design tool for studying the effect of design parameters on the resulting system response. 
Other uses of this approach include fault detection from sensor signals in a wide variety of engineering operations.  
\end{abstract}
%%%%%%%%%%%%%%%%%%%%%%%%%%%%%%%%%%%%%%%%%%%%%%%%%%%%%%%%%%%%%%%%%%%%%%
%!TEX root = ..\MSP_main_IDETC.tex
%-------------------------------
%*******************************
\section{INTRODUCTION} \label{sec:intro}
The progress in modern manufacturing operations has given designers unprecedented freedom in conceptualizing structural and machine components. 
However, one concern in going from a design idea to a functional product is whether the manufactured component or assembly will behave unexpectedly under certain operation conditions. 
In order to address this concern, it is necessary to either write predictive models and simulate different scenarios, or manufacture prototypes and collect field data. 
If the design parameter space is large, it is possible to acquire a large amount of data that cannot be manually analyzed by human operators. 
Therefore, it is necessary to develop and utilize analysis tools that can autonomously analyze the resulting data and classify the resulting system behavior, for example, as erratic/chaotic, or regular/periodic. 
These tools can also prove useful in the context of fault detection after the part or the assembly are put into operation to guard against unexpected failures using sensory signals in the form of time series. 

In this work we describe an approach for characterizing a time series as chaotic or periodic based on the structure of its complex network embedding. 
The embedding of the time series into a network or a graph is achieved using ordinal partitions \cite{McCullough2015}. 
Embedding is often necessary because the true underlying model of the data is unknown, and all that is available usually is a time-indexed, observed quantity such as acceleration or temperature. 
Therefore, embedding can help identify how the dynamics of the system evolve starting with a one dimensional recording of data.  
Currently, perhaps the most popular tool for time series analysis is Takens' embedding \cite{Takens1981}, where a time series is embedded into an $n$-dimensional Euclidean space using a uniform subsampling by some constant delay $\tau$. 
Both the embedding dimension $n$ and the delay $\tau$ are parameters that need to be selected using, for example, false nearest neighbors approach~\cite{Kennel1992} and the first minimum of the mutual information function~\cite{Fraser1986}, respectively. 
When the chosen parameters lead to a successful Takens' embedding, the reconstructed state space can qualitatively be used to infer the underlying dynamics of the system that generated the time series. 
The above approach for time series embedding assumes a deterministic time series or one with little additive noise. Additionally, the resulting embedding does not take into account the order of the points in the time series. 
Practically, Takens' embedding is often used with small $n$, typically less than seven or eight, although visualizing any embeddings with $n>3$ is difficult or impossible.  

Alternatively, the time series can be embedded into a graph $G=(E,V)$ where $V$ are the vertices of the graph and $E$ are its edges. 
In contrast to Takens' continuous representation of the data, Graph embedding leads to a discrete representation which can be more easily visualized even for high-dimensional data. 
The idea is that the shape of the graph can provide information on the underlying structure of the dynamical system. 

A commonly used network embedding based on phase space reconstruction is the recurrence network~\cite{Donner2010}. 
The nodes in a recurrence network are formed by each of the embedded vectors, e.g., from from Takens' embedding. 
An edge is added between two nodes if the Euclidean distance between the corresponding embedded vectors is less than a user-specified threshold $\epsilon \in [0, \infty)$. 
Although recurrence networks embedding can represent the underlying structure of the phase space, it also introduces a variable $\epsilon$ that needs to be selected.  
Khor et al.~used a $k$-nearest neighbor network~\cite{Khor2016} where $\epsilon$ was replaced with another user parameter $k \in \mathbb{Z}^+$. 
Another option for network embedding that we mention here is the visibility graph~\cite{Lacasa2008,Luque2009}. 
However, recently, a network embedding approach based on ordinal partitions was described \cite{McCullough2015}. 

This method does not require a distance threshold like recurrence networks, but rather forms a networks based on the permutation transitions within a time series. 
In section~\ref{ssec:per_homology} below a description of the basic idea of this approach is provided in more detail, which is the method we chose to work with in this manuscript. 

\begin{figure}[h] 
    \centering
    \includegraphics[scale = 0.43]{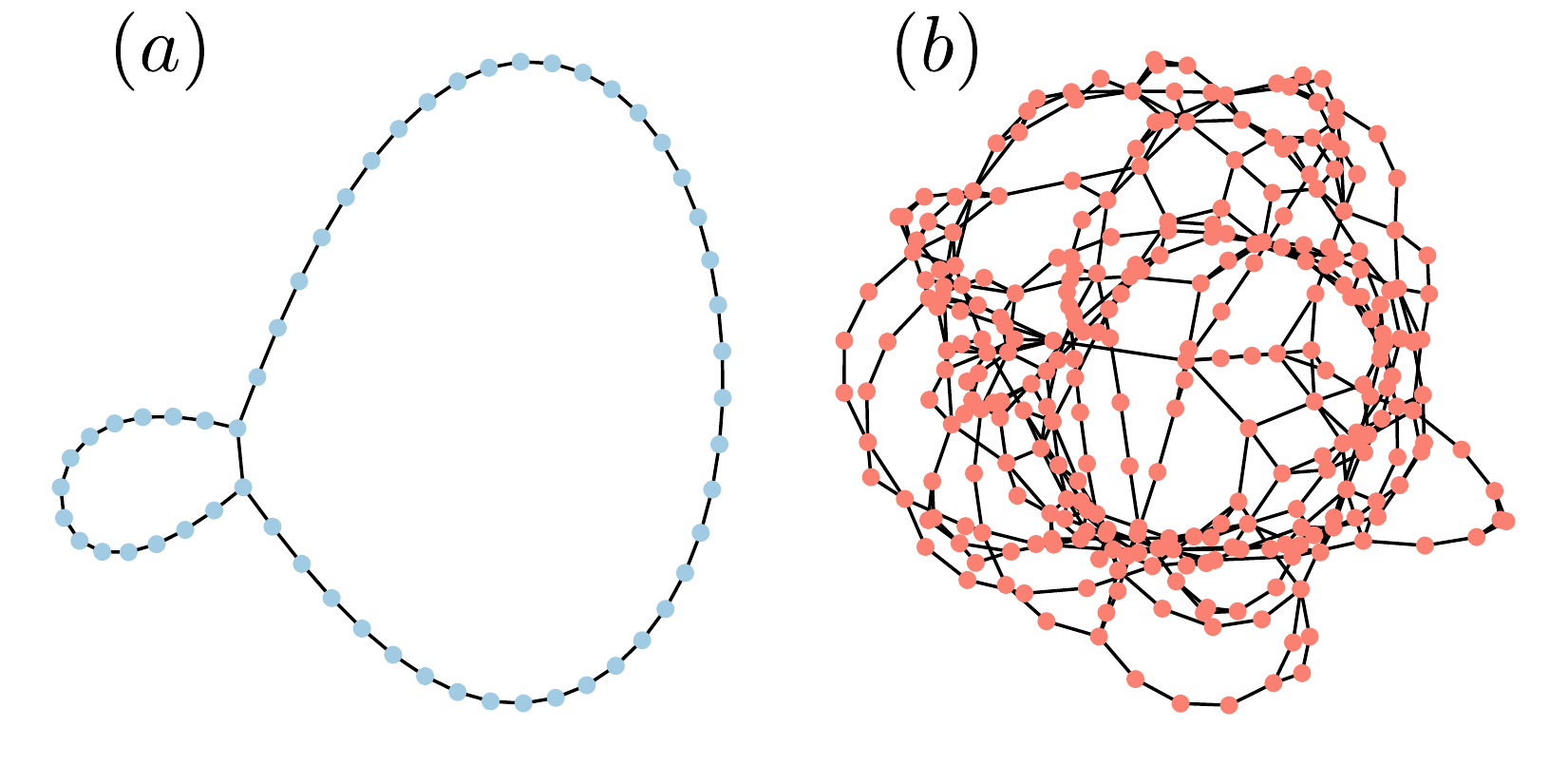}
    \caption{Example ordinal partition networks generated from a (a) periodic time series and (b) chaotic time series.}
    \label{fig:chaotic_vs_periodic_figure}
\end{figure}

Ordinal partitions embedding provides a framework for embedding the time series into a graph; however, the challenge becomes in identifying the system state (periodic or chaotic) using the structure or shape of the resulting graph. 
More specifically, we would like to have a quantitative measure to classify the visual differences we see in ordinal partition embeddings of periodic and chaotic signals, see Fig.~\ref{fig:chaotic_vs_periodic_figure} for an example.
%--------------------------------------------------
%--------------------------------------------------

One tool that has been shown to successfully quantify the shape of ordinal partitions network comes from Topological Data Analysis (TDA). 
Specifically, in~\cite{Myers2019}, persistent homology, the flagship tool from TDA was successfully used to distinguish between periodic and chaotic signals. 
However, previous work on the ordinal partition and TDA pipeline only considered simulated signals of autonomous nonlinear systems; therefore, extrapolating the validity of the findings to physical systems, especially with time dependent forcing, has not been previously investigated. 

In this work, we study the applicability of the method described in~\cite{Myers2019} to detect transitions from periodic to chaotic dynamics in a magnetic, single pendulum under base excitation,see Fig.~\ref{fig:MSP_rendering_and_model} and Section~\ref{sec:experiment}. 
This physical system provides rich, non-linear dynamics over a wide range of parameter values and demonstrates both periodic and chaotic dynamics. 
Further, it ties in with other applications where similar systems are used such as as energy harvesting~\cite{Castagnetti2019, Li2014} and mass dampers~\cite{Lian2018}. 

The paper begins by introducing the experimental setup and model in Section~\ref{sec:experiment}. A quick introduction to ordinal partition networks and persistent homology is then provided in Section~\ref{sec:background}. 
In Section~\ref{ssec:point_summaries} we introduce point summaries for quantifying the shape of the networks down to a single statistic. 
Then, in Section~\ref{sec:method}, an example response from the experimental magnetic pendulum with relatively complex periodic dynamics is used to demonstrate the functionality of the method for analyzing the time series through the shape of the resulting networks. 
Finally, in Section~\ref{sec:results} we provide results for the variation of the base excitation amplitude to detect dynamic state transitions over a wide variety of dynamic responses with varying complexity. 

\section{EXPERIMENT: THE MAGNETIC PENDULUM} \label{sec:experiment}
\textbf{Note}: a Computer Aided Design (CAD) model and design document for the pendulum used for the experimental section of this manuscipt is available through \textit{GitHub} at \textit{https://github.com/Khasawneh-Lab/simple\_pendulum}.

The driven magnetic pendulum is a well known system to exhibit chaos~\cite{Siahmakoun1997, Tran2013, Khomeriki2016}. 
Therefore, we designed and built a magnetic pendulum apparatus, and utilized the ordinal partition embedding and TDA to characterize the dynamics of the resulting signals. 

In this section we derive a simplified equation of motion using Lagrange's approach. 
The design, manufacturing, and equipment used for the experiment are also explained.  
Additionally, we describe our methods for estimating and measuring the constants that appear in the equation of motion.

\subsection{MODEL}
\begin{figure}[h] 
    \centering
    \includegraphics[scale = 0.22]{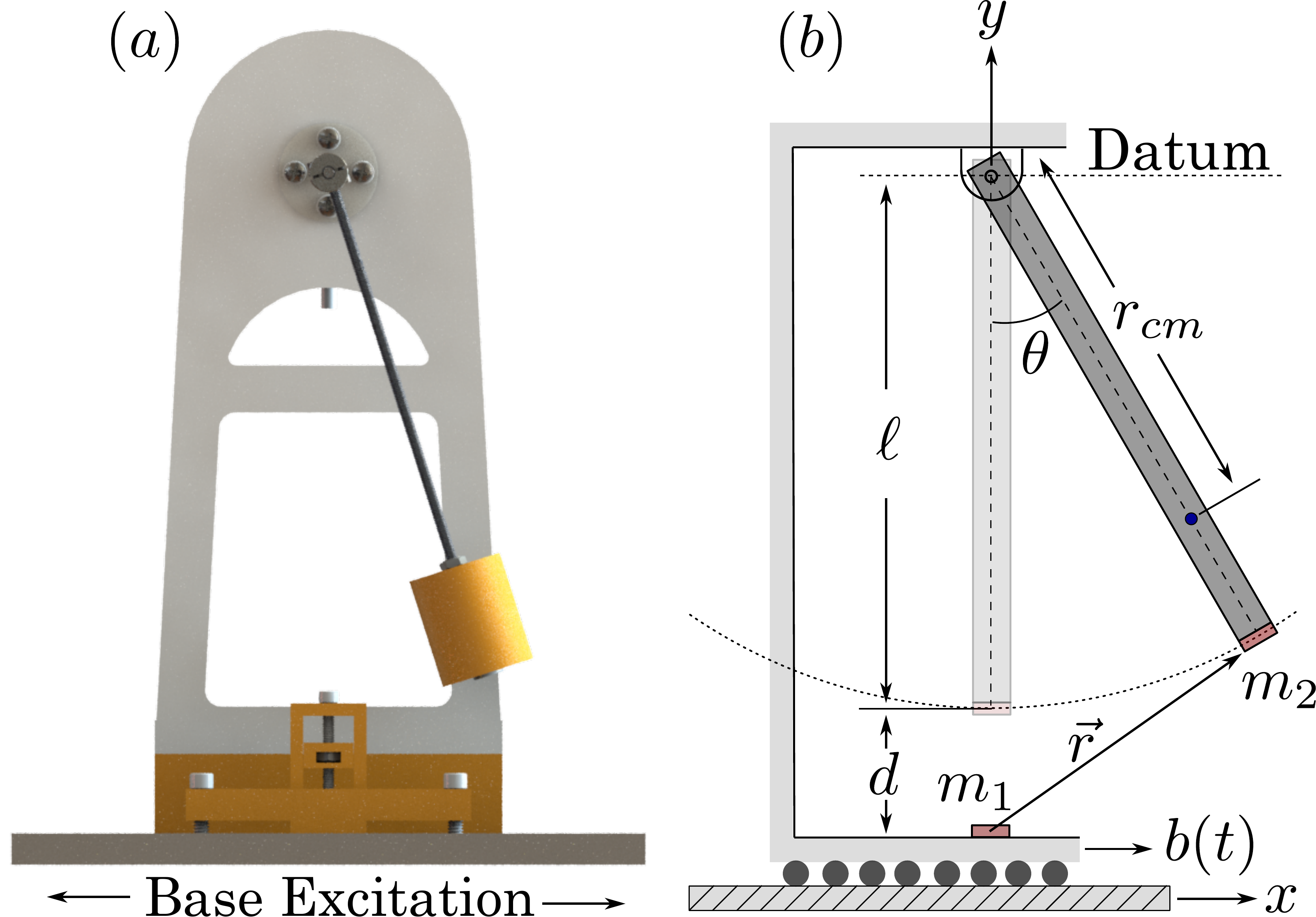}
    \caption{Rendering of experimental setup in comparison to reduced model, where $b(t) = A\sin(\omega t)$ is the base excitation with frequency $\omega$ and amplitude $A$, $r_{cm}$ is the effective center of mass of the pendulum, $d$ is the minimum distance between magnets $m_1 = m_2 = m$ (modeled as dipoles), and $\ell$ is the length of the pendulum.}
    \label{fig:MSP_rendering_and_model}
\end{figure}

We begin by deriving the equations of motion for the physical system shown in Fig.~\ref{fig:MSP_rendering_and_model}.  
Let the total mass of the rotating components be $M$, the distance from the rotation center $O$ to the mass center of the rotating assembly $r_{\rm cm}$, and the mass moment of inertia of the rotating components about their mass center be $I_{\rm cm}$. 
Further, assume that the magnetic interactions are well approximated by a dipole model with $m_1 = m_2 = m$ representing the magnitudes of the dipole moment. 
%--------------------------------------------------
%--------------------------------------------------
To develop the equation of motion, we use Lagrange's equation (Eq.~\eqref{eq:Lagrange}), so the potential energy $V$, kinetic energy $T$, and non-conservative moments $R$ are needed. 
In this analysis the damping moments and the moments generated from the magnetic interaction are treated as non-conservative. 
The potential and kinetic energy are defined as
\begin{equation}
\begin{split}
T & = \frac{1}{2}M|{\vec{v}}_{cm}|^2 + \frac{1}{2}I_{cm}\dot{\theta}^2, \\
V & = -M g {r}_{cm}\cos(\theta),
\end{split}
\label{eq:TV}
\end{equation}
where $\vec{v}_{cm}$ is the velocity of the mass center given by
\begin{equation}
\vec{v}_{cm}  = r_{cm}\dot{\theta}\left[ \cos(\theta)\hat{\epsilon}_x + \sin(\theta)\hat{\epsilon}_y\right] + A\cos(\omega t)\hat{\epsilon}_x.
\label{eq:v_cm}
\end{equation}

In Eq.~\eqref{eq:v_cm}, $A\cos(\omega t)$ is introduced from the base excitation $b(t) = A\cos(\omega t)$ in the $x$ direction with $A$ as the amplitude and $\omega$ as the frequency and $\hat{\epsilon}_x$ and $\hat{\epsilon}_y$ are the unit vectors in the $x$ and $y$ directions, respectively.

The non-conservative moments are caused by the energy lost to damping. 
For our analysis, we consider three possible mechanisms of energy dissipation: Coulomb damping $\tau_c$, viscous damping $\tau_v$, and quadratic damping $\tau_q$. 
We chose to use all three mechanisms of damping due to previous work on damping estimation for a pendulum similar to the one we used~\cite{Petrushenko2017}. 
These three moments are defined as
\begin{equation}
\begin{split}
\tau_c & = \mu_c {\rm sgn}(\dot{\theta}), \\
\tau_v & = \mu_v \dot{\theta}, \\
\tau_q & = \mu_q \dot{\theta}^2 {\rm sgn}(\dot{\theta}), 
\end{split}
\label{eq:damping}
\end{equation}
where $\mu_c$, $\mu_v$, and $\mu_q$ are the coefficient for Coulomb, viscous, and quadratic damping, respectively. 

\begin{figure}[h] 
    \centering
    \includegraphics[scale = 0.22]{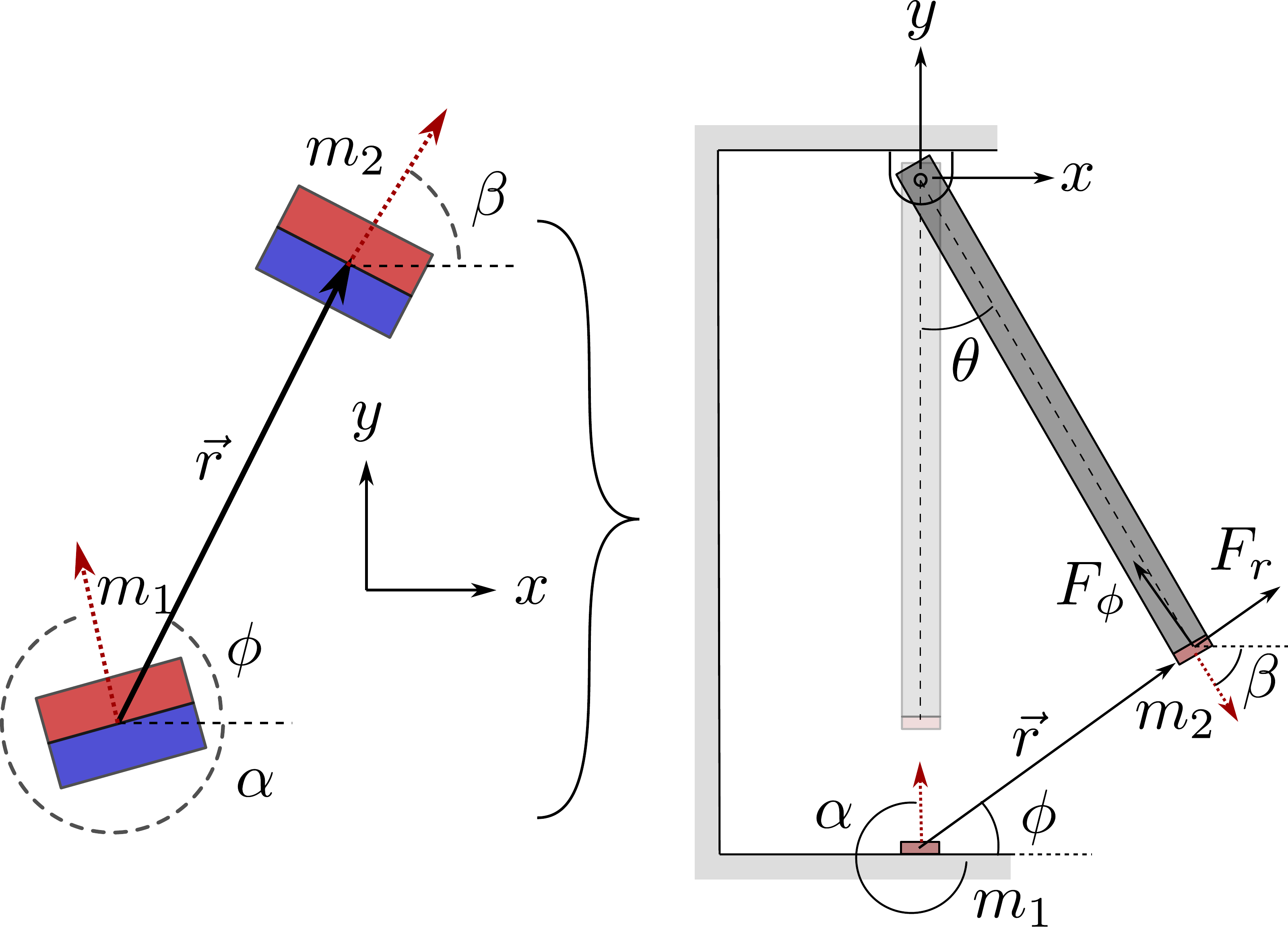}
    \caption{A comparison between a generic, in-plane magnetic model in global coordinates and the equivalent magnetic forces in the pendulum model $F_r$ and $F_\phi$ (see Eq.~\eqref{eq:F_magnetic}).}
    \label{fig:MSP_magnetic_force_diagram}
\end{figure}

To begin the derivation of the torque induced from the magnetic interaction $\tau_m$, consider two, in-plane magnets as shown on the left side of Fig.~\ref{fig:MSP_magnetic_force_diagram}. 
The red side of the magnet in the figure represents its north-pole.
From this representation, the magnetic force acting on each magnet is calculated as
\begin{equation}
\begin{split}
F_r & = \frac{3\mu_o m^2}{4 \pi r^4} \left[ 2c(\phi-\alpha) c(\phi-\beta) - s(\phi-\alpha)s(\phi-\beta)\right], \\
F_\phi & = \frac{3\mu_o m^2}{4 \pi r^4} \left[s(2\phi-\alpha-\beta)\right],
\end{split}
\label{eq:F_magnetic}
\end{equation}
where $m_1$ and $m_2$ are the magnetic moments, $\mu_o$ is the magnetic permeability of free space, and $c(*) = \sin(*)$ and $s(*) = \sin(*)$. 
Equation~\eqref{eq:F_magnetic} assumes that the cylindrical magnets used in the experiment can be approximated as a dipole. 
We later show that this assumption is satisfactory in Fig.~\ref{fig:magnetic_force_experiment} of Section~\ref{ssec:parameters}.
These magnetic forces are then adapted to the physical pendulum as shown on the right side of Fig.~\ref{fig:MSP_magnetic_force_diagram}, with $\alpha = \pi/2$ and $\beta = \pi/2 - \theta$. 
Additionally, $\phi$ and $r$ are calculated from $\theta$, $d$, and $\ell$ from Fig.~\ref{fig:MSP_rendering_and_model} as 
\begin{align}
\label{eq:phi}
\phi &= \frac{\pi}{2} - \arcsin\left(\frac{\ell}{r}\sin(\theta)\right), \quad \text{ and }\\
\label{eq:r}
r &= \sqrt{{[\ell\sin(\theta)]}^2 + {[d + \ell(1-\cos(\theta))]}^2}.
\end{align}
The moment induced by the magnetic interaction is then
\begin{equation}
\tau_m  = \ell F_r \cos(\phi-\theta) - \ell F_\phi \sin(\phi-\theta).
\label{eq:tau_m}
\end{equation}

Using $\tau_m$ from Eq.~\eqref{eq:tau_m} and the non-conservative torques from Eq.~\eqref{eq:damping}, $R$ is defined as
\begin{equation}
R = \tau_c + \tau_v + \tau_q + \tau_m.
\label{eq:R}
\end{equation}
Finally, the equation of motion for the base-excited magnetic single pendulum is found by substituting the above expressions into Lagrange's equation and noting that $L=T-V$
\begin{equation}
\frac{\partial}{\partial t}\left(\frac{\partial L}{\partial \dot{\theta}}\right) - \frac{\partial L}{\partial \theta} + R = 0. 
\label{eq:Lagrange}
\end{equation}

Equation~\eqref{eq:Lagrange} was symbolically manipulated to express it in state space format using Python's \textit{Sympy} package. 
Then, the system was simulated at a frequency of $f_s = 60$ Hz using Python's \textit{odeint} function from the \textit{Scipy} library. 

\subsection{EQUIPMENT AND EXPERIMENTAL DESIGN}
\begin{figure}[h] 
    \centering
    \includegraphics[scale = 0.22]{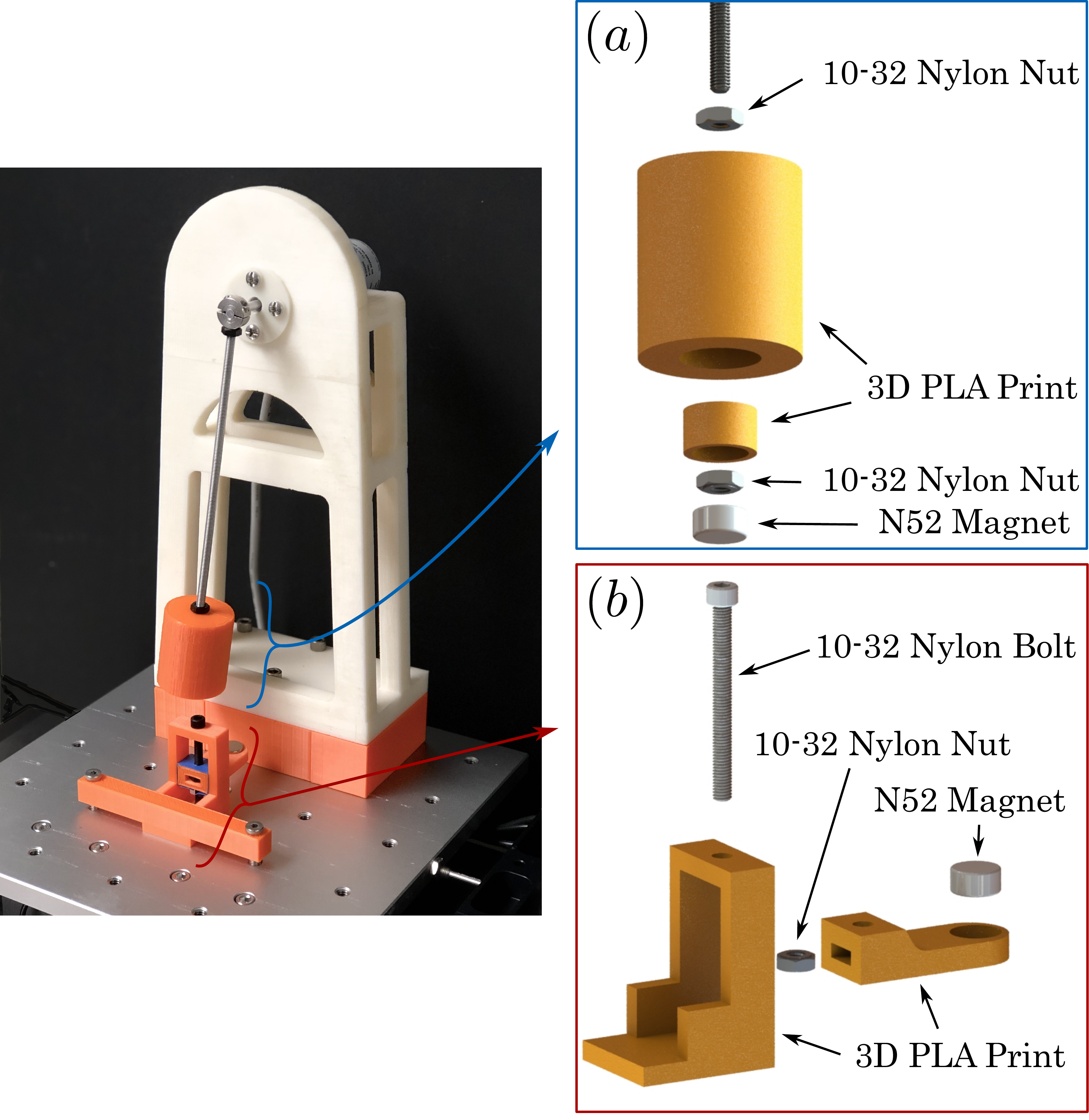}
    \caption{Manufacturing overview with experimental setup. In Fig.~(a), an exploded view of the end mass (100$\%$ infill 3D printed PLA components) is shown with the magnet press fit into end of pendulum. In Fig.~(b), an exploded view of the linear stage controlling the vertical position of the lower magnet.}
    \label{fig:assembly_process}
\end{figure}

The setup of the experiment was manufactured by extending the capabilities of a previously manufactured simple pendulum~\cite{Petrushenko2017}. To increase the non-linearity, in-plane magnets on the base as well as at the end of the pendulum were added. 
To assume a permeability of free space $\mu_0$, any ferromagnetic material within the vicinity was removed, which made the use of 3D printed components critical. 
In Fig.~\ref{fig:assembly_process} an overview of the utilized, 3D-printed components are shown. 
Specifically, Figs.~\ref{fig:assembly_process}~(a)~and~(b) show exploded views of the end mass of the pendulum, and the linear stage for controlling the distance $d$, respectively. 
The magnets used are two, approximately identical, rare-earth (neodymium) N52 permanent magnets with a radius and length of 6.35 mm (1/4"). 

Table~\ref{tab:equipment} provides a list of the item, description, and manufacturer for all of the experimental equipment used to collect the rotational data from the magnetic single pendulum under base excitation.
\begin{table}[h]
\centering
\begin{tabular}{ccc}
\hline
Item & Description & \multicolumn{1}{l}{Manufacturer} \\ \hline
Shaker & 113 Electro-Seis & APS \\
DC Power Supply & Model 1761 & BK Precision \\
Accelerometer & Model 352C22 & Piezotronics \\
Rotary Encoder & UCD-AC005-0413 & Posital \\
Data Acquisition & USB-6356 & Nat. Inst. \\
PC & OptiPlex 7050 & Dell \\
\hline
\end{tabular}
\caption{Equipment used for experimental data collection.} \label{tab:equipment}
\end{table}

\subsection{PHYSICAL PARAMETERS AND CONSTANTS} \label{ssec:parameters}
\begin{figure}[h] 
    \centering
    \includegraphics[scale = 0.53]{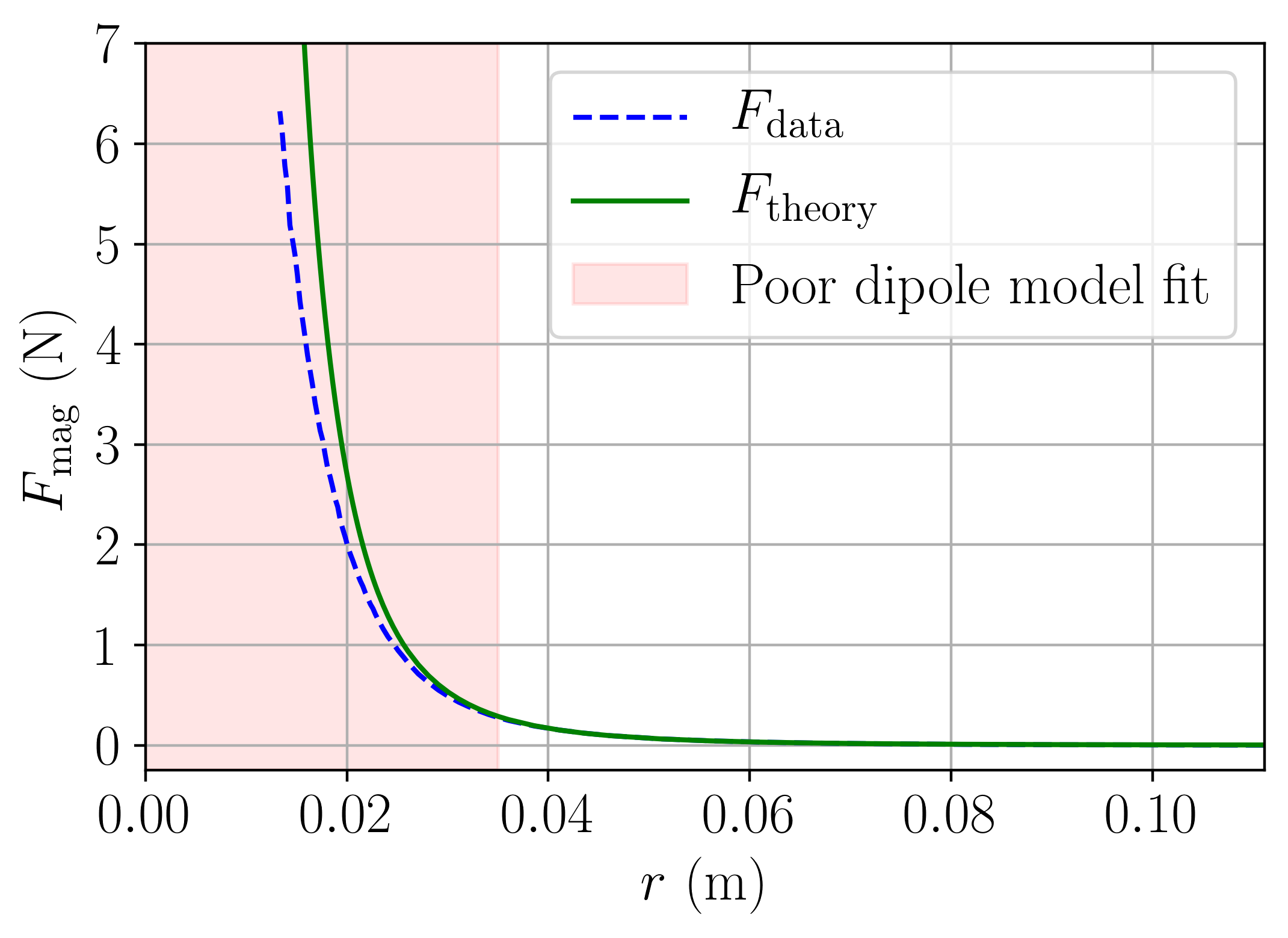}
    \caption{Measured repulsion force as a function of distance compared to theoretical force in Eq.~\eqref{eq:F_magnetic} with $\theta = 0$. The theoretical force $F_{\rm theory}$ is based on dipole model with a dipole moment $m = 0.85$ cm, which was estimated using a curve fit to the region where the magnetic thickness $T \ll r$. Region of poor fit is marked for $r<0.035$ m.}
    \label{fig:magnetic_force_experiment}
\end{figure}

To estimate the magnetic dipole moment $m$ of the cylindrical magnets used (see Fig.~\ref{fig:assembly_process}), we performed an experiment similar to the one described in~\cite{Gonzalez2016}. 
When the distance between the magnets is less than a critical value $r_c$, modeling the magnets as dipoles can lead to large errors since the dipole model does not accurately approximate the repulsive force between the magnets. 
This distance was estimated as $r_c = 0.035$ m (see Fig.~\ref{fig:magnetic_force_experiment}). 
Additionally, in the region where $r>r_c$, the force curve, a function of scale $r^{-4}$, was fit to the curve to estimate the magnetic dipole moment as $m=0.85$ Cm.

The other parameter values as well as their uncertainties (when applicable) are provided in Table~\ref{tab:parameters}, which are in reference to Fig.~\ref{fig:MSP_rendering_and_model}. 
Most of these parameters were either estimated using \textit{SolidWorks} or by multiple direct measurements. 
\begin{table}[h]
\centering
\begin{tabular}{lll}
\hline
Parameter (units) & Value & Uncertainty ($\pm \sigma$) \\ \hline
$d$ (m) & 0.36 & 0.005 \\
$\ell$ (m) & 0.208 & 0.005 \\
$g$ ($\rm m/s^2$) & 9.81 & - \\
$M$ (kg) & 0.1038 & 0.005 \\
$r_{\rm cm}$ (m) & 0.188 & - \\
$\omega$ (rad/s) & $3\pi$ & - \\
$\mu_0$ (Cm) & $1.257\times 10^{-6}$ & - \\
$m$ (Cm) & 0.85 & - \\
$\mu_c$ (-) & 0.002540 & 0.000020 \\
$\mu_v$ (-) & 0.000015 & 0.000003 \\
$\mu_q$ (-) & 0.000151 & 0.000020 \\
\hline
\end{tabular}
\caption{Equation of motion parameters to simulated pendulum with associated uncertainty.} \label{tab:parameters}
\end{table}

\begin{figure}[h] 
    \centering
    \includegraphics[scale = 0.46]{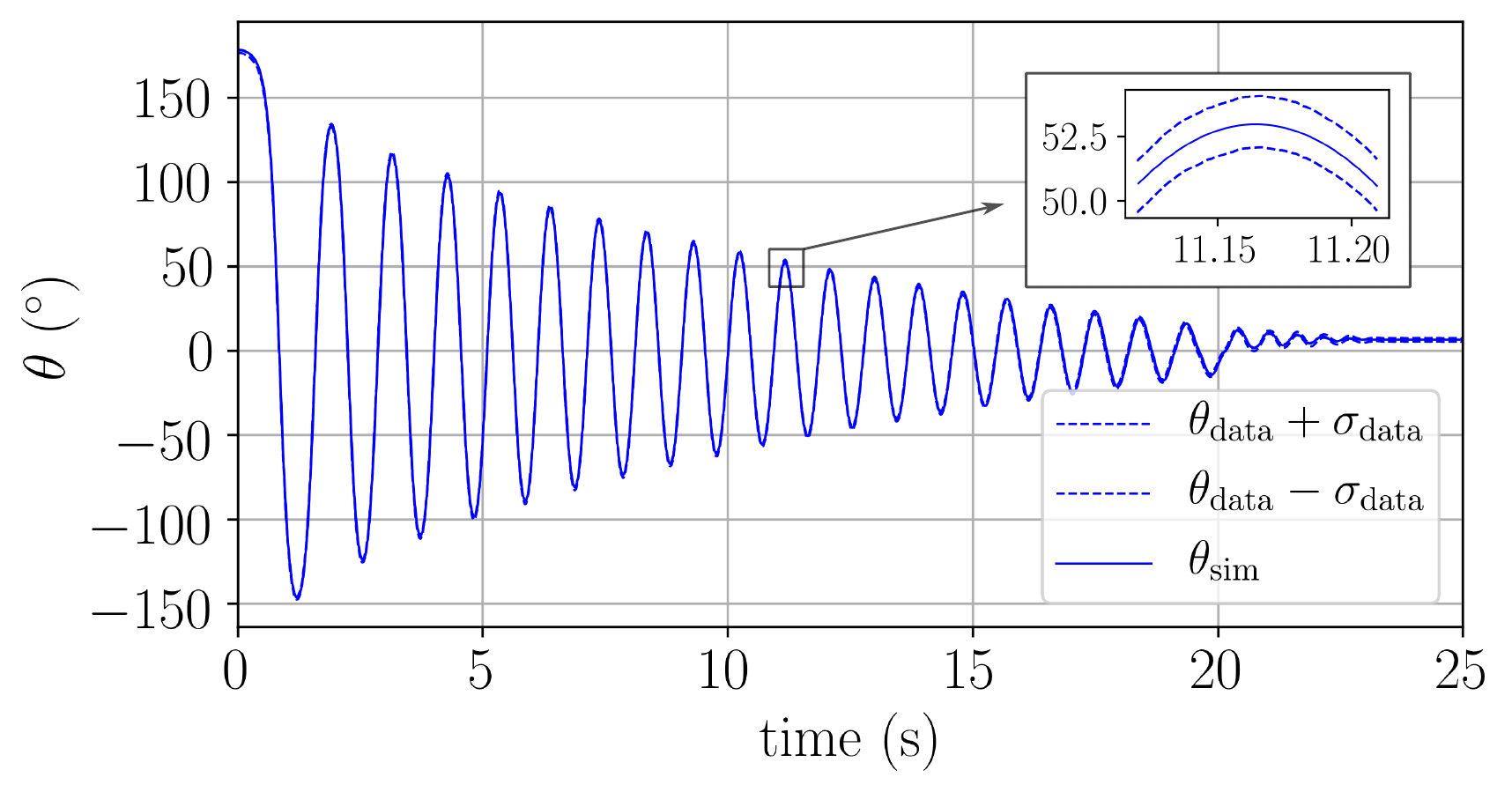}
    \caption{Free drop test between collect angular position data $\theta_{\rm data}$ with encoder uncertainty $\sigma_{\rm data}$ and the simulated response $\theta_{\rm sim}$. As shown in the zoomed-in region, the simulated response is within the bounds of uncertainty of the actual response.}
    \label{fig:free_drop_comp}
\end{figure}

To validate the parameters, an experiment and simulation of a free drop of the pendulum are compared. The resulting angle $\theta(t)$ is shown in Fig.~\ref{fig:free_drop_comp}, which shows a very similar response between simulation and experiment. Additionally, the simulation is within the bounds of uncertainty of the encoder $\sigma_{\rm data} = 1^{\circ}$ as shown in the zoomed in region of Fig.~\ref{fig:free_drop_comp}. 
%%%%%%%%%%%%%%%%%%%%%%%%%%%%%%%%%%%%%%%%%%%%%%%%%%%%%%%%%%%%%%%%%%%%%%
%!TEX root = ..\arxiv_MSP_manuscript.tex
%-------------------------------
% %*******************************
\section{BACKGROUND}
\label{sec:background}
%*******************************
This section introduces the tools needed to form ordinal partition networks as well as analyze their shape using persistent homology.
\subsection{ORDINAL PARTITION NETWORKS}
\begin{figure*}[t] 
    \centering
    \includegraphics[scale = 0.57]{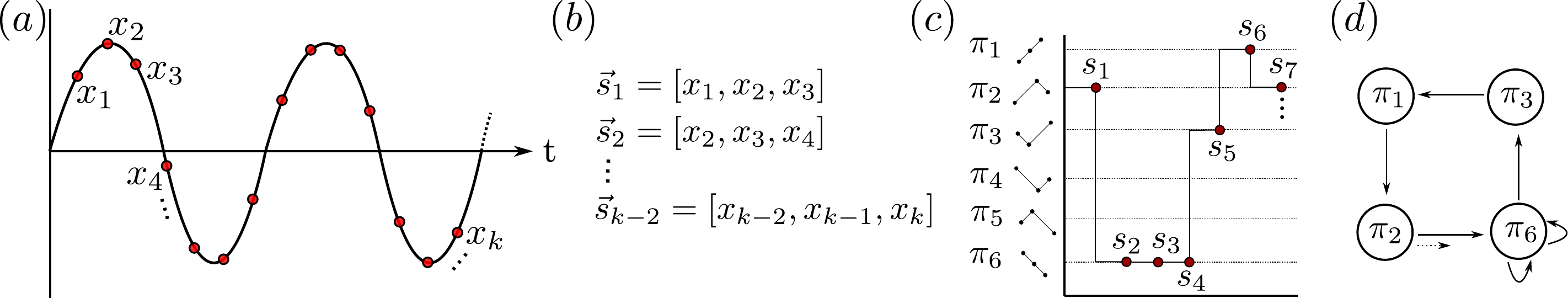}
    \caption{Example ordinal partition network embedding of a time series ${\bf{x}} = x_1, x_2, x_3, x_4, \ldots, x_k$ shown in Fig.~(a). 
    In Fig.~(b), subsamples $\vec{s}_i$ of the time series of dimension $n=3$ are obtained using a uniform subsampling $\tau$. 
    In Fig.~(c) the ordinal transformation of the vectors $\vec{s}_i$ into their corresponding permutation $\pi_j$ with $j \in [1,6]$ is shown. 
    Figure (d) records the permutations and their transitions in the form of a network where $\pi_2 \rightarrow \pi_6 \rightarrow \pi_5 \rightarrow \pi_1 \rightarrow \pi_2 \rightarrow \ldots \rightarrow \pi_4$.}
    \label{fig:network_formation_overview}
\end{figure*}
Ordinal partitions embedding uses the permutation transitions within the time series~\cite{McCullough2015,Zhang2017,Ruan2019}. 
Permutations as a time series analysis tool were first popularized by Bandt and Pompe through permutation entropy~\cite{Bandt2002}. 
However, this relatively simple statistical summary of entropy does not capture any information about the time ordering of the permutation transitions. 
A natural way to then capture the frequency and ordering of the permutation transitions is through a complex network or graph. 
For ordinal partition networks, the vertices $V$ are the collection of permutations found within the time series $x(t)$ and the edges $E$ are formed from the transitions between permutations. 
To elucidate how these networks are obtained, consider the example shown in Fig.~\ref{fig:network_formation_overview}. 
The example begins with the simple time series in Fig.~\ref{fig:network_formation_overview}-(a), which is defined as $x(t) = \sin(t)$ and was sampled at a rate of $f_s \approx 7$ Hz. 
The sampled data points are ordered as ${\bf{x}} = x_1, x_2, x_3, x_4, \ldots, x_k$, where $k$ is the total number of samples. 
In order to define the permutations, we need to set two parameters: a delay $\tau$ and the dimension $n$. 
The delay parameter $\tau$ represents a uniform subsampling of the time series, while the dimension $n$ determines the size and the possible number of the used permutations. 
Specifically, a dimension $n$ means that there is a total of $n!$ possible permutations. 
Note that these two terms are synonymous to the ones used in Takens' embedding; however, there is not yet a theory that connects the two sets of terms.  
% ${\bf{x}}$ is then embedded into a state space through a strategy known as Taken's embedding~\cite{Takens1981}. 
% Typically,  Taken's embedding is used to reconstruct the phase space of an attractor through vector embedding. 
% However, permutations summarize this phase space through a binning based on the ordinal partitioning of the values from the embedded vectors. 
% The delay embedding vectors from Taken's theorem are defined as $\vec{s}_i = [x_i, x_{i+\tau}, x_{i+2\tau},\ldots, x_{i+(n-1)\tau}]$, where $n$ is the embedding dimension and $\tau$ is the embedding delay.

We select $\tau$ and $n$ using multi-scale permutation entropy as suggested in~\cite{Myers2019a,Riedl2013}. 
The formation of these vectors is shown for our example time series in Fig.~\ref{fig:network_formation_overview}-(b), where $n=3$ and $\tau=1$. Let us now consider $\vec{s}_1 = [x_1, x_2, x_3]$ to demonstrate the permutation assignment. First, an ordinal ranking of  $\vec{s}_1$ results in $x_1<x_3<x_2$. This forms the permutation of form $\pi_2$ from $s_1$ as shown in Fig.~\ref{fig:network_formation_overview}-(c). Continuing this permutation assignment for the rest of the embedded vectors yields the repeating sequence of permutations $\pi_2 \rightarrow \pi_6 \rightarrow \pi_5 \rightarrow \pi_1 \rightarrow \pi_2 \ldots$ (ignoring repeated permutations) as shown in Fig.~\ref{fig:network_formation_overview}-(c). This permutation sequence can be represented as a network (see fig~\ref{fig:network_formation_overview}-(d)) where edges are formed from the permutation transitions with the nodes being the visited permutations. It should also be mentioned that for permutations of dimension $n$, there are $n!$ possible permutations. Therefor, as the dimension increase, the complexity of the time series is better captured. However, with exceedingly high dimensions ($n>8$), the computation time becomes increasingly large.

\subsection{PERSISTENT HOMOLOGY OF NETWORKS} \label{ssec:per_homology}
This section briefly describes computing the persistent homology of complex networks; a more detailed description is provided in~\cite{Myers2019}. 
It begins by introducing simplicial complexes and their filtration, followed homology groups and persistent homology. 
 % (a tool from TDA), which we can summarize as the tracking of the formation and collapse of homology groups through a filtration. 

\paragraph{SIMPLICIAL COMPLEXES:}
Simplicial complexes are one of the backbones of persistent homology.  
A simplicial complex $K$ is a collection of simplices, where a simplex $\sigma$ is a collection of vertices from the full set of vertices $V$ such that $\sigma \subseteq V$. 
In our application the vertices that compose the simplicies are the vertices of the graph itself. 
The dimension $d$ of a simplex is based on the number of vertices in the simplex or as ${\rm dim}(\sigma) = |\sigma|-1$. 
From this notation, a point is $d=0$,  an edge is $d=1$, a face is $d=2$, and so on.

\begin{figure}[h] 
    \centering
    \includegraphics[scale = 0.6]{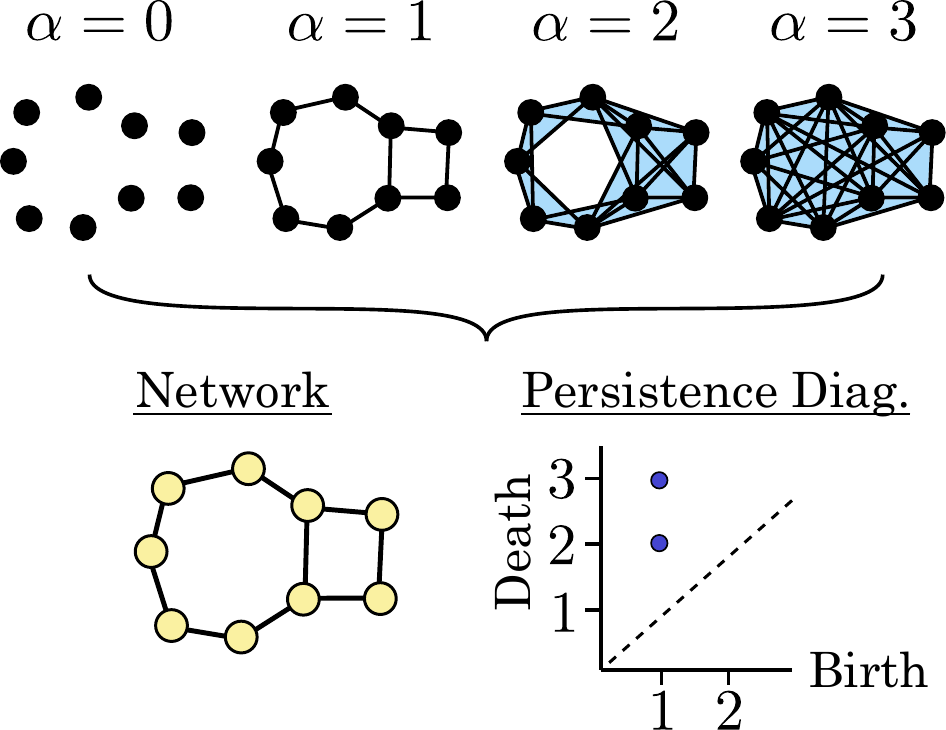}
    \caption{An example filtration of $\alpha \in [0,1, 2, 3]$ showing the nodes when $\alpha = 0$, both loop structures being born at $\alpha = 1$ (this also returns us to our original network), the death of the smaller of the loops at $\alpha = 2$, and the death of the larger loop at $\alpha = 3$. This filtration and the associated births and deaths are recorded in a persistence diagram, which summarizes the loops in $\mathbb{R}^2$ with the coordinates of a feature as $(\alpha_{\rm birth}, \alpha_{\rm death})$.}
    \label{fig:pers_diag_from_network}
\end{figure}

A filtration is a collection of simplicial complexes such that $K_1 \subseteq K_2 \subseteq K_3,\ldots, K_N$, where each simplex is generated at a specific filtration level. 
Filtrations are usually accomplished by incrementing a threshold parameter $\alpha$ that increases the number of connected vertices, thus forming more simplices and growing the overall simplicial complex.  
To illustrate a filtration for our application, Fig.~\ref{fig:pers_diag_from_network} shows a simple network example with 9 nodes as shown in the bottom left. At the top of the figure we show the filtration over multiple scales of $\alpha$. 
The vertices are connected, i.e., and edge is added, when the shortest (unweighted) distance between them is less or equal to the scale $\alpha$ or $d(u,v) \leq \alpha$, where $u$ and $v$ are two vertices in the graph and $d(*)$ is the shortest path distance. 

\paragraph{HOMOLOGY:}
A homology group can be geometrically understood as simple structures of dimension $d$, where a point is a $d=0$ structure, a loop is $d=1$, and a void is $d=2$. In this work we will only use loops ($d=1$) for analyzing the networks. 

\paragraph{PERSISTENT HOMOLOGY:} 
The main idea of persistent homology is to track the formation and collapse of certain homology groups throughout the filtration of the simplicial complex $K$. We can think of the formation of a feature at a filtration level $\alpha_B$ as its birth and the collapse at a filtration level $\alpha_D$ as its death. 
The lifetime $L$ of a feature is then calculated as $L = \alpha_D-\alpha_B$. 
Let us now return to our simple network example in Fig.~\ref{fig:pers_diag_from_network}. The bottom right of the figure shows the persistence diagram, which is used to track the births $\alpha_B$ and deaths $\alpha_D$ of the $d=1$ homology groups through the coordinate $(\alpha_B, \alpha_D)$. 
At a filtration level of $\alpha = 0$ we do not have any loops that have formed, but rather just the original vertices of the network. 
However, both of the loops are born at $\alpha = 1$. 
At the next filtration level, $\alpha = 2$, the smaller of the two loops dies, which is tracked in the persistence diagram as the point $(1,2)$. 
Then, at the final filtration level $\alpha = 3$, our larger loop also dies, which is again recorded in the persistence diagram as the point $(1,3)$. 
We can then calculate the lifetimes by taking the difference between the death and birth filtration levels of the two loops for lifetimes of $1$ and $2$ for the small and big loop, respectively.

\subsection{POINT SUMMARIES OF THE PERSISTENCE DIAGRAM} \label{ssec:point_summaries}
Our next goal is to develop summary statistics of the resulting persistence diagrams of the unweighted and undirected ordinal partition networks. This will be done through two statistics: the periodicity score and persistent entropy.

\paragraph{PERIODICITY SCORE:}
The first summary statistic we develop is the periodicity score, which summarizes how periodic a network is based on a comparison to an unweighted cycle graph $G'$ with $n$ vertices. 
If we are using the distance metric of the shortest path with an unweighted graph, then all loops will form at $\alpha_B = 1$ and they will fill in at $\alpha_D = \lceil \tfrac{n}{3} \rceil$.
This results in the persistence diagram $D'$ from $G'$ with exactly one point with a lifetime of
\begin{equation}
    L_n = {\rm maxpers}(D') = \left\lceil \frac{n}{3} \right\rceil-1 .
\end{equation}
Let us now assume we are given another unweighted graph from our ordinal partition network $G$ with $n$ vertices. 
This results in the persistence diagram $D$, where the maximum lifetime of $D$ is used to calculate the network's periodicity score as
\begin{equation}
    P(D) = 1-\frac{{\rm maxpers}(D)}{L_n}.
    \label{eq:Rn_Equation}
\end{equation}
This peridoicity score is similar in nature to that developed in \cite{Perea2015}, but applied to unweighted networks. 
Additionally, it is normalized in such a way that $P(D) \in [0,1]$, with $P(D) = 0$ only if $G$ is a cycle graph.
%-------------------------------------------

\paragraph{NORMALIZED PERSISTENT ENTROPY:}
Persistent entropy was first developed by Chintakunta et al.~\cite{Chintakunta2015} as an implementation of the original definition of information entropy by Shannon~\cite{Shannon1948}. 
Persistent entropy is calculated as the entropy of the lifetimes from a persistence diagram.
This summary statistic is defined as
\begin{equation}
E(D) = - \sum_{x \in D} \frac{{\rm pers}(x)}{\mathscr{L}(D)}\log_2\left(\frac{{\rm pers}(x)}{\mathscr{L}(D)}\right),
\label{eq:pers_ent}
\end{equation}
where $\mathscr{L}(D) = \sum_{x \in D} {\rm pers}(x)$ is the sum of lifetimes of points in the diagram.
To make it possible to make comparisons across multiple persistence diagrams, we normalize $E$ according to
\begin{equation}
E'(D) = \frac{E(D)}{\log_2\big(\mathscr{L}(D))}.
\label{eq:persent_norm}
\end{equation}
%%%%%%%%%%%%%%%%%%%%%%%%%%%%%%%%%%%%%%%%%%%%%%%%%%%%%%%%%%%%%%%%%%%%%%
%!TEX root = ..\MSP_main_IDETC.tex
%-------------------------------
%*******************************
\section{METHOD}
\label{sec:method}
\begin{figure}[h] 
    \centering
    \includegraphics[scale = 0.5]{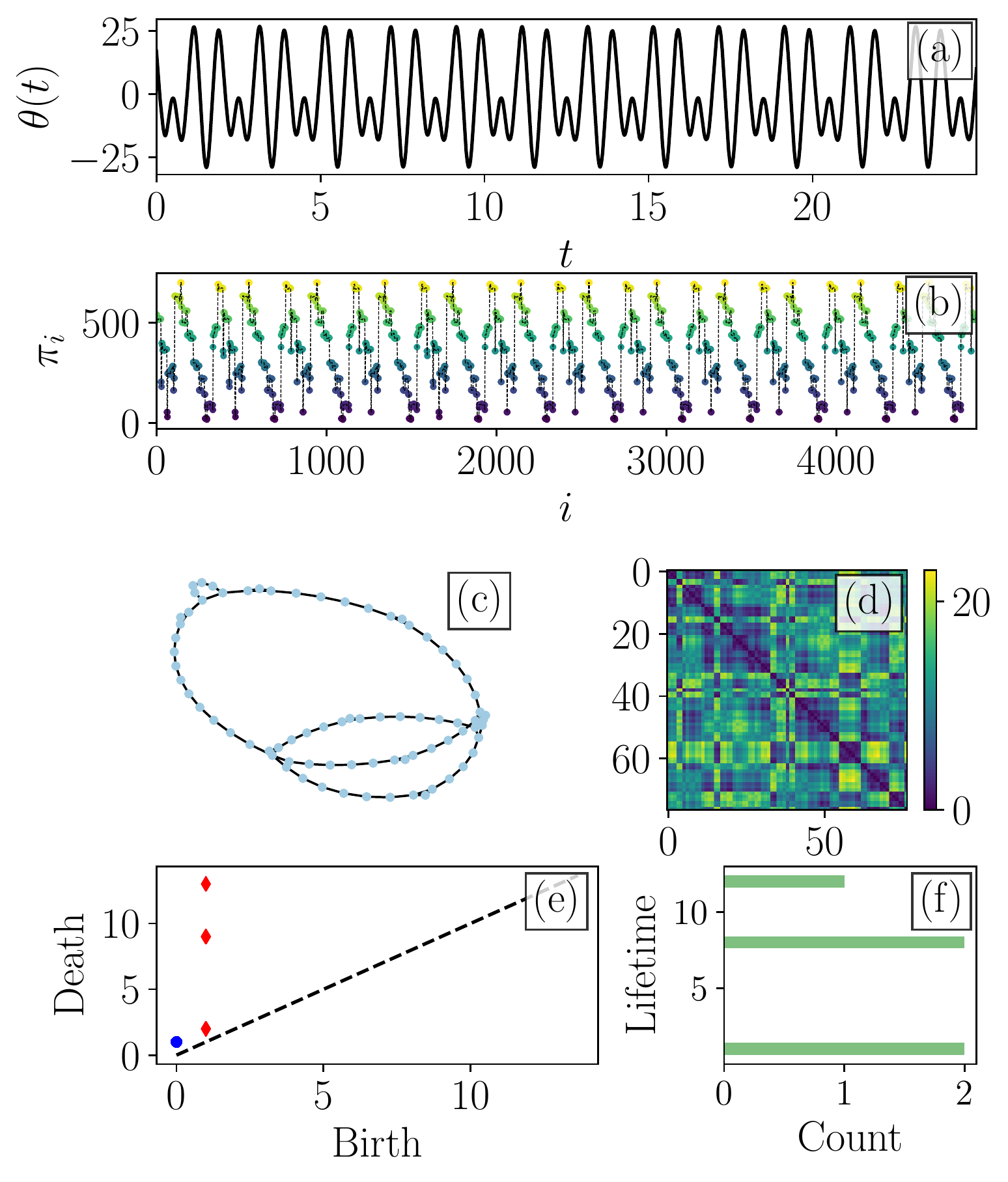}
    \caption{Example of method applied to experimental data with a periodic response Fig.~(a). In Fig.~(b) the sequence of permutations are shown for $n=6$ with the associated ordinal partition network in Fig.~(c). In Fig.~(d) the distance matrix (using an unweighted network and short path distance) is shown, which was used to compute a persistence diagram with multiplicity shown in Fig.~(e)~and~(f), respectively.}
    \label{fig:periodicOP_network_process_MSP}
\end{figure}

To demonstrate the method, we will be using a time series obtained from the angular position $\theta(t)$ of the magnetic pendulum experiment shown in Fig.~\ref{fig:MSP_rendering_and_model} with base excitation amplitude $A = 0.08$ m and frequency $\omega = 1.5$ Hz. 
This forcing amplitude results in the periodic time series shown in Fig.~\ref{fig:periodicOP_network_process_MSP}-(a). 
The resulting permutation sequence as well as the unweighted, undirected network are shown in Figs.~\ref{fig:periodicOP_network_process_MSP}-(b)~and~(c), respectively. The network exhibits a rather simple structure with one large loop, two smaller loops, and two insignificantly small loops. 
The distance between nodes is shown through a shortest-path distance matrix (see Fig.~\ref{fig:periodicOP_network_process_MSP}-(d)). 
With the distance matrix known, the persistence diagram is obtained as described in Section~\ref{ssec:per_homology}, which summarizes the loops as 1-D features with lifetimes of $[12, 8, 8, 1, 1]$. 
Additionally, a histogram is used to show the lifetime multiplicity, i.e., how many points are overlaid in each location of the persistence diagram. 
The periodicity score was calculated as $P(D) \approx 0.61$ and the persistent entropy was calculated as $E'(D) \approx 0.45$ using the lifetimes in Fig.~\ref{fig:periodicOP_network_process_MSP}-(f).

\begin{figure}[h] 
    \centering
    \includegraphics[scale = 0.5]{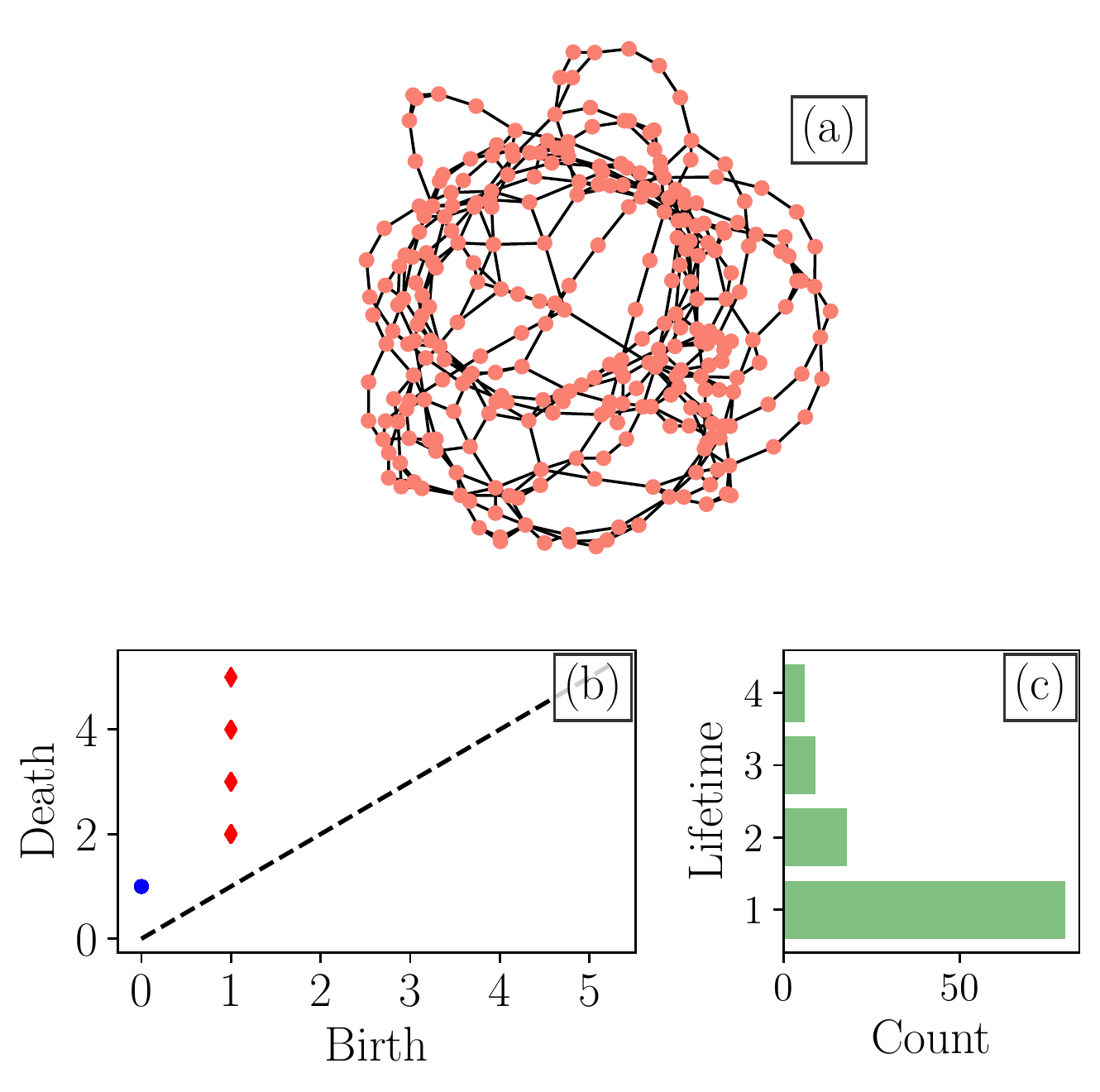}
    \caption{Example of method applied to experimental data with a chaotic response Fig.~(a). In Fig.~(b) the sequence of permutations are shown for $n=6$ with the associated ordinal partition network in Fig.~(c). In Fig.~(d) the distance matrix (using an unweighted network and short path distance) is shown, which was used to compute a persistence diagram with multiplicity shown in Fig.~(e)~and~(f), respectively.}
    \label{fig:chaotic_OP_network_process_MSP_networkPD}
\end{figure}

To make a fair comparison, the same process as shown in Fig.~\ref{fig:periodicOP_network_process_MSP} is applied to a time series generated from a base excitation with $A = 0.085$ and frequency $\omega = 1.5 Hz$, which results in a chaotic response. 
The resulting network from the permutation sequence is shown in Fig.~\ref{fig:chaotic_OP_network_process_MSP_networkPD}-(a). 
It is clear that the network from the chaotic time series shows significantly more loops with, in general, smaller loop sizes. 
The size and quantity of these loops are shown in the persistence diagram of the network with the lifetimes (with multiplicity) shown in Fig.~\ref{fig:chaotic_OP_network_process_MSP_networkPD}-(b) and (c), respectively. 
The periodicity score was calculated as $P(D) \approx 0.95$ and the persistent entropy was calculated as $E'(D) \approx 0.90$.

These examples show how persistent homology of complex networks can be used to detect a change in complexity of the time series. 
We will now use the point summaries to detect the state transitions when varying the amplitude $A$ of the base excitation.

%%%%%%%%%%%%%%%%%%%%%%%%%%%%%%%%%%%%%%%%%%%%%%%%%%%%%%%%%%%%%%%%%%%%%%
%!TEX root = ..\MSP_main_IDETC.tex
%-------------------------------
%*******************************
\section{RESULTS}
\label{sec:results}
%*******************************
\begin{figure}[h] 
    \centering
    \includegraphics[scale = 0.45]{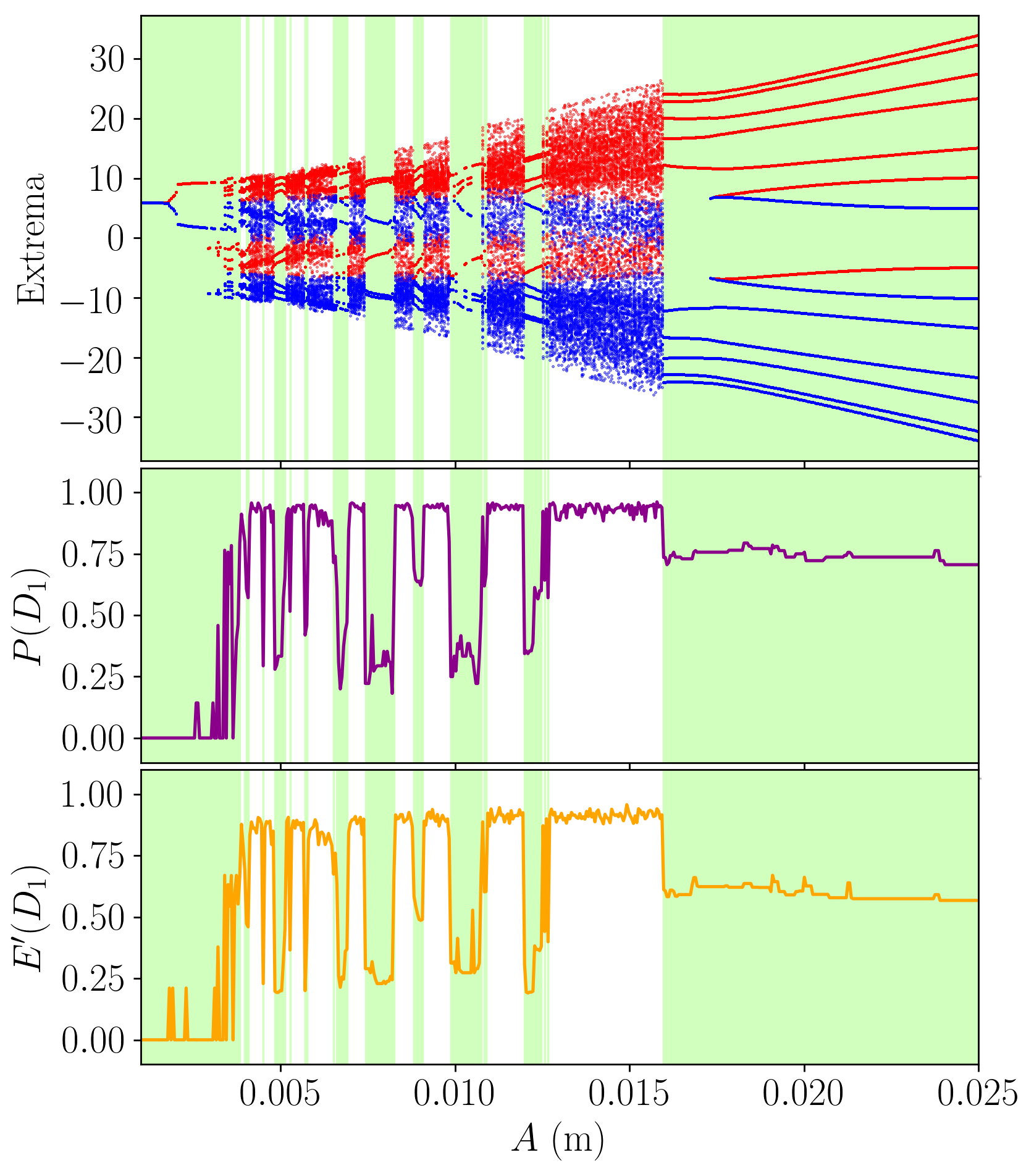}
    \caption{Bifurcation analysis of magnetic pendulum through numeric simulations with variation of base excitation amplitude $A \in [0.001, 0.025]$ meters with a step size of $5 \times 10^{-5}$ meters. The top figure shows the local extrema in the generated time series with red being maxima and blue being minima. The middle and lower figures shows the periodicity score and normalized persistent entropy, respectively. Additionally, regions where both $P(D)<0.8$ and $E'(D)<0.8$ are shaded in green, which represents a periodic time series.}
    \label{fig:OP_networks_A_bifurcation}
\end{figure}
To show that the persistence diagrams from ordinal partition networks can distinguish periodic from chaotic time series over a wide range of parameters, we use a bifurcation analysis. 
This was done by simulating the magnetic pendulum with the variation of the base excitation amplitude $A$ from 0.001 meters to 0.025 m by increments of $5 \times 10^{-5}$ m with the frequency held constant at $\omega = 3\pi$ rad/s. 
At each amplitude, a time series was simulated for 400 seconds at a sampling frequency of 200 Hz, where only the last 100 seconds were used to avoid the transient response. 
The maxima and minima from each time series were recorded as a qualitative method for detecting dynamic state changes. 
These extrema are shown in the top figure of Fig.~\ref{fig:OP_networks_A_bifurcation} with the red and blue data points representing maxima and minima, respectively. 
Moreover, regions with a periodic response are highlighted (light green). 
In addition to the local extrema, the periodicity score $P(D_1)$ and persistent entropy $E'(D_1)$ were calculated from the resulting persistence diagrams of the ordinal partition networks. 
These two scores show distinct drops for the periodic regions and values near 1 for chaotic regions. 
Specifically, the periodic response can be separated from the chaotic one by setting the thresholds 0.80 and 0.75 for $P(D_1)$ and $E'(D_1)$, respectively. Additionally, for the periodic responses, the complexity is captured by the value of each score. 
%

%%%%%%%%%%%%%%%%%%%%%%%%%%%%%%%%%%%%%%%%%%%%%%%%%%%%%%%%%%%%%%%%%%%%%%
%!TEX root = ..\MSP_main_IDETC.tex
%-------------------------------
%*******************************
\section{CONCLUSION}
\label{sec:conclusion}
%*******************************
This paper described a novel method for analyzing a time series from a mechanical system through ordinal partition networks and TDA. 
The example that we designed and built to experimentally validate the developed approach is a magnetic simple pendulum with base excitation. 
In addition to the experimental model, we also derive the governing differential equation and fit the corresponding parameters. 
The generated time series for the physical experiment and the numerical simulation of the model were then analyzed using ordinal partition networks and persistent homology from TDA. 
This was done by showing how a time series can be transformed into an ordinal partition networks, which captures a summary of the phase space reconstruction through the permutation transitions. 
As shown in Fig.~\ref{fig:chaotic_vs_periodic_figure}, ordinal partition networks of periodic time series result in a relatively simple structure, while those from a chaotic response have an irregular shape. 
To summarize these shape differences we used the persistence diagram from persistent homology. 
Specifically, we computed two summary statistics: the periodicity score and persistence entropy.

Our results for the experimental data showed that the persistent homology of ordinal partition networks is suitable for analyzing real world data. Specifically, in Fig.~\ref{fig:periodicOP_network_process_MSP} and Fig.~\ref{fig:chaotic_OP_network_process_MSP_networkPD} we showed that there is a definite structural differences between the ordinal partition networks for the periodic and chaotic responses, even with the inherent noise present in the system. Additionally, we were able to successfully distinguish between periodic and chaotic responses using persistent homology and the summary statistics for a relatively short time series response making this method suitable for real world applications.

By simulating the governing differential equations of the pendulum over a wide range of the base excitation amplitude, Fig.~\ref{fig:OP_networks_A_bifurcation} showed that the summary statistics clearly distinguished between periodic and chaotic responses. 
This is evidenced by clear dips in the summary statistics in the periodic regions versus their higher values in the chaotic ones. 
Specifically, a threshold of approximately 0.80 for the periodicity score $P(D_1) $and 0.75 for persistent entropy $E'(D_1)$ successfully separated periodic from chaotic time series with a value above the threshold signifying a chaotic response. 
We remark that in contrast to the work in \cite{Myers2019}, this paper represents the first application of the described approach to time series obtained from non-autonomous systems. 
Therefore, we believe that one of its strengths in contrast to the delay-reconstruction approach is that it does not require any special embedding procedures for forced systems, see \cite{Stark1999,Stark2001}. 

%%%%%%%%%%%%%%%%%%%%%%%%%%%%%%%%%%%%%%%%%%%%%%%%%%%%%%%%%%%%%%%%%%%%%%
\section*{ACKNOWLEDGMENTS}
\label{sec:acknowledgments}
%*******************************
This material is based upon work supported by the National Science Foundation under grant
nos. CMMI-1759823 and DMS-1759824 with PI FAK.

\bibliographystyle{plain}
\bibliography{msp_bib}

\end{document}